\documentclass[aps, prd, a4paper, 10pt, showpacs, superscriptaddress, floatfix, nofootinbib, reprint]{revtex4-1}
\tighten
\usepackage{color}
\usepackage{xcolor}
\usepackage{amsmath,amssymb,graphicx}
\usepackage{wasysym}
\usepackage{bm,lipsum}
\usepackage{times}
\usepackage{microtype}
\usepackage[utf8]{inputenc}
\usepackage{multirow}

\usepackage{booktabs}
\usepackage{subfigure}
\usepackage[normalem]{ulem}
\usepackage[varg]{txfonts}
\usepackage{bm,graphics,graphicx,epsfig,color,ulem}
\usepackage[colorlinks, pdfborder={0 0 0}]{hyperref}
\definecolor{LinkColor}{rgb}{0.75, 0, 0}
\definecolor{CiteColor}{rgb}{0, 0.5, 0.5}
\definecolor{UrlColor}{rgb}{0, 0, 0.75}
\hypersetup{linkcolor=LinkColor}
\hypersetup{citecolor=CiteColor}
\hypersetup{urlcolor=UrlColor}
\allowdisplaybreaks
\newcommand{\red}{\color{red}}

\def\msun{{\rm M}_{\odot}}
\newcommand\code[1]{\textsc{\MakeLowercase{#1}}}

\def\gsim{\mathrel{
\rlap{\raise 0.511ex \hbox{$>$}}{\lower 0.511ex
\hbox{$\sim$}}}}

\def\lsim{\mathrel{
\rlap{\raise 0.511ex \hbox{$<$}}{\lower 0.511ex
\hbox{$\sim$}}}}

\begin{document}

\title{Towards inference of overlapping gravitational wave signals}
\author{Elia Pizzati} 
\affiliation{Institute for Gravitation and the Cosmos, Department of Physics, Penn State University, 
University Park PA 16802, USA}
\affiliation{Scuola Normale Superiore, Piazza dei Cavalieri 7, 56126 Pisa, Italy}
\author{Surabhi Sachdev} 
\affiliation{Institute for Gravitation and the Cosmos, Department of Physics, Penn State University, 
University Park PA 16802, USA}
\affiliation{Leonard E. Parker Center for Gravitation, Cosmology, and Astrophysics, University of Wisconsin-Milwaukee, Milwaukee, WI 53201, USA}
\author{Anuradha Gupta} 
\affiliation{Department of Physics and Astronomy, The University of Mississippi, University, Mississippi 38677, USA}
\author{B. S. Sathyaprakash} 
\affiliation{Institute for Gravitation and the Cosmos, Department of Physics, Penn State University, 
University Park PA 16802, USA}
\affiliation{Department of Astronomy and Astrophysics, Penn State University, University Park PA 16802, USA}
\affiliation{School of Physics and Astronomy, Cardiff University, Cardiff, CF24 3AA, United Kingdom}
\date{\today}

\begin{abstract}

Merger rates of binary black holes, binary neutron stars, and neutron star-black hole binaries in the local Universe (i.e., redshift $z=0$), inferred from the Laser Interferometer Gravitational Wave Observatory (LIGO) and Virgo, are 16--130 Gpc$^{-3}$~yr$^{-1}$,  13--1900 Gpc$^{-3}$~yr$^{-1}$, and 7.4--320 Gpc$^{-3}$ yr$^{-1}$,  respectively. These rates suggest that there is a significant chance that two or more of these signals will overlap with each other during their lifetime in the sensitivity-band of future gravitational-wave detectors such as the Cosmic Explorer and Einstein Telescope. The detection pipelines provide the coalescence time of each signal with an accuracy $\mathcal{O}(10\,\rm ms)$. We show that using a prior on the coalescence time from a detection pipeline, it is possible to correctly infer the properties of these {\it overlapping signals} with the current data-analysis infrastructure. We study different configurations of two overlapping signals created by non-spinning binaries, varying their time and phase at coalescence, as well as their signal-to-noise ratios. We conclude that, for the scenarios considered in this work, parameter inference is robust provided that their coalescence times in the detector frame are more than $\sim 1$--2$\mathrm{s}$. Signals whose coalescence epochs lie within $\sim 0.5\,\rm s$ of each other suffer from significant biases in parameter inference, and new strategies and algorithms would be required to overcome such biases.
\end{abstract}

\maketitle

\section{Introduction}

The advent of the third generation (3G) gravitational-wave (GW) observatories, such as the Cosmic Explorer (CE) \cite{Evans:2021gyd, 2019arXiv190704833R, Reitze:2019dyk} and the Einstein Telescope (ET) \cite{Punturo:2010zz}, will offer the possibility to observe binary coalescence events from redshifts $z\sim$~10--50,  thanks to an order of magnitude improved strain and frequency sensitivity compared to the current generation of detectors of Advanced LIGO \cite{LIGOScientific:2014pky}, Advanced Virgo \cite{VIRGO:2014yos}, and KAGRA \cite{KAGRA:2018plz}. 
 Indeed, 3G observatories will have unprecedented sensitivity to detect coalescence events from an epoch when the Universe was still in its infancy assembling its first stars and will routinely detect mergers with stupendously large signal-to-noise ratios of several thousands \cite{Sathyaprakash:2012jk, Vitale:2016icu, Maggiore:2019uih, Evans:2021gyd}. An order of magnitude greater redshift reach and access to extremely high-fidelity signals compared to current interferometers promises many new discoveries, while allowing completely independent, precision tests of cosmological models, alternative gravity theories, and astrophysical scenarios of compact binary formation and evolution \cite{Evans:2021gyd, Maggiore:2019uih}. With an expected rate of hundreds of thousands of binary coalescence signals each year \cite{Baibhav:2019gxm, Sachdev:2020bkk, Maggiore:2019uih, Evans:2021gyd} on top of weak, but persistent, radiation from isolated neutron stars \cite{Sathyaprakash:2012jk}, rare bursts from supernova and other transient sources and stochastic backgrounds \cite{Regimbau:2016ike}, 3G observatories demand novel algorithms for signal detection and characterization. Therefore, a proper understanding of systematics arising from overlapping loud and quiet signals alike will answer a range of scientific questions that are at the forefront of fundamental physics and astronomy, as well as a realistic estimation of the computational cost. 
 
According to current estimates, 3G observatories are expected to detect hundreds of thousands of binary black hole (BBH) and binary neutron star (BNS) mergers each year \cite{Baibhav:2019gxm, Sachdev:2020bkk, Maggiore:2019uih, Evans:2021gyd}. If we take account of the fact that signals will last longer due to a lower starting frequency ($3\,\mathrm{Hz}$ for ET and $5\,\mathrm{Hz}$ for CE), then it is clear that 3G data will be dominated by many overlapping signals \cite{Regimbau:2012ir, Meacher:2015rex, Regimbau:2016ike, Samajdar:2021egv, Relton:2021cax}. The problem of overlapping signals producing a confusion background in future terrestrial detectors was identified more than a decade ago \cite{Regimbau:2009rk}.  The problem poses two challenges: first, the detection of individual signals could, in principle, be affected by the presence of multiple signals. Second, the current Bayesian inference methods \cite{Veitch:2014wba, Ashton:2018jfp} may not guarantee unbiased estimation of source parameters, which is crucial to deliver the science promises of 3G observatories.

A similar issue has been tackled, in a different context, by the LISA (Laser Interferometer Space Antenna) community. LISA is expected to produce a data set containing many overlapping astrophysical signals: galactic white dwarf binaries are persistent sources of gravitational waves and they produce a ``foreground'' noise \cite{Crowder:2004ca} that could masquerade the detection and parameter estimation of other astrophysical signals. Several authors have studied the problem of both detection \cite{Cornish:2006ms, Littenberg:2011zg, MockLISADataChallengeTaskForce:2009wir} and Bayesian inference \cite{Cornish:2005qw, Crowder:2006eu} in this context, while others have focused on searching for the global solution to the full family of potential signals \cite{Littenberg:2020bxy, Robson:2017ayy, Petiteau:2012zq}. A parallel effort has been made by other studies \cite{Cornish:2014kda, Chatziioannou:2021ezd, Cornish:2020dwh} to characterize the overlapping between GW signals and glitches in the context of LIGO/Virgo data analysis. These studies represent a useful reference that could guide the development of new algorithms specifically suited to deal with the parameter estimation of multiple signals in the context of terrestrial detectors. 

However, no effort to study the problem of inference in the case of 3G terrestrial detectors has so far been made. Given the relevance of this specific problem, an exploratory study of the capabilities of current parameter estimation methods in the context of overlapping signals in terrestrial detectors appears to be necessary. With this consideration in mind, we aim to characterize the conditions for which parameter estimation is possible with the current algorithms for overlapping signals and to identify regions in the signal parameter space that create significant biases in the inference process, for which novel algorithms would be required.

Detecting overlapping GW signals has been shown to be possible by two ET mock data challenges \cite{Regimbau:2012ir, Meacher:2015rex}. These studies were able to correctly identify and recover signals even when they were overlapping with multiple others.  
Even though the signal detection may provide unbiased results, however, there is no guarantee that the parameter inference in the case of overlapping signals is possible within the current framework. This is because current methods heavily rely on the efficiency of sampling algorithms, which are used to explore the posterior distribution of parameters. If we analyze overlapping signals with the current parameter estimation (PE) procedures (i.e., the assumption that the parameter space for multiple signals is the same as in the case of data containing only one signal at a time), we expect Markov Chains and the posterior distribution to exhibit a non-trivial behavior such as slowly or non-convergence of chains, multi-modal and biased posterior distributions, etc. 

To this end, we deploy the Fisher information matrix formalism to gauge the limit between the region where overlapping signals could lead to biases in parameter inference and the region where they don't. The Fisher study tells us that as long as the difference in the merger time $\Delta t_C$ of two overlapping signals is larger than the accuracy $\delta t$ with which their merger times can be measured (i.e., $\Delta t_C \gg \delta t$), irrespective of how long the individual signals are, parameter inference will not cause significant biases. We exploit this result in the Bayesian analysis of mock data by choosing the prior on the merger epoch as determined by the signal detection pipelines, which is about $\delta t_C \sim \mathcal{O}(10 \,{\rm ms})$ \cite{liting2014}. Indeed, most signals are recovered by search pipelines with a bias of $\delta t_C <20$~ms. A conservative prior on the merger time could be a factor of 10 to 20 larger or at most 500 ms. Thus, two overlapping signals with their merger times separated by larger than about $\approx 2\,\mathrm{s}$ are not expected to suffer from any systematic biases. Hence, it suffices to consider the extent to which overlapping signals pose a problem for $\Delta t_C \lesssim 2 \rm\, s.$

\begin{figure*}
    \centering
    \includegraphics[width=0.70\textwidth]{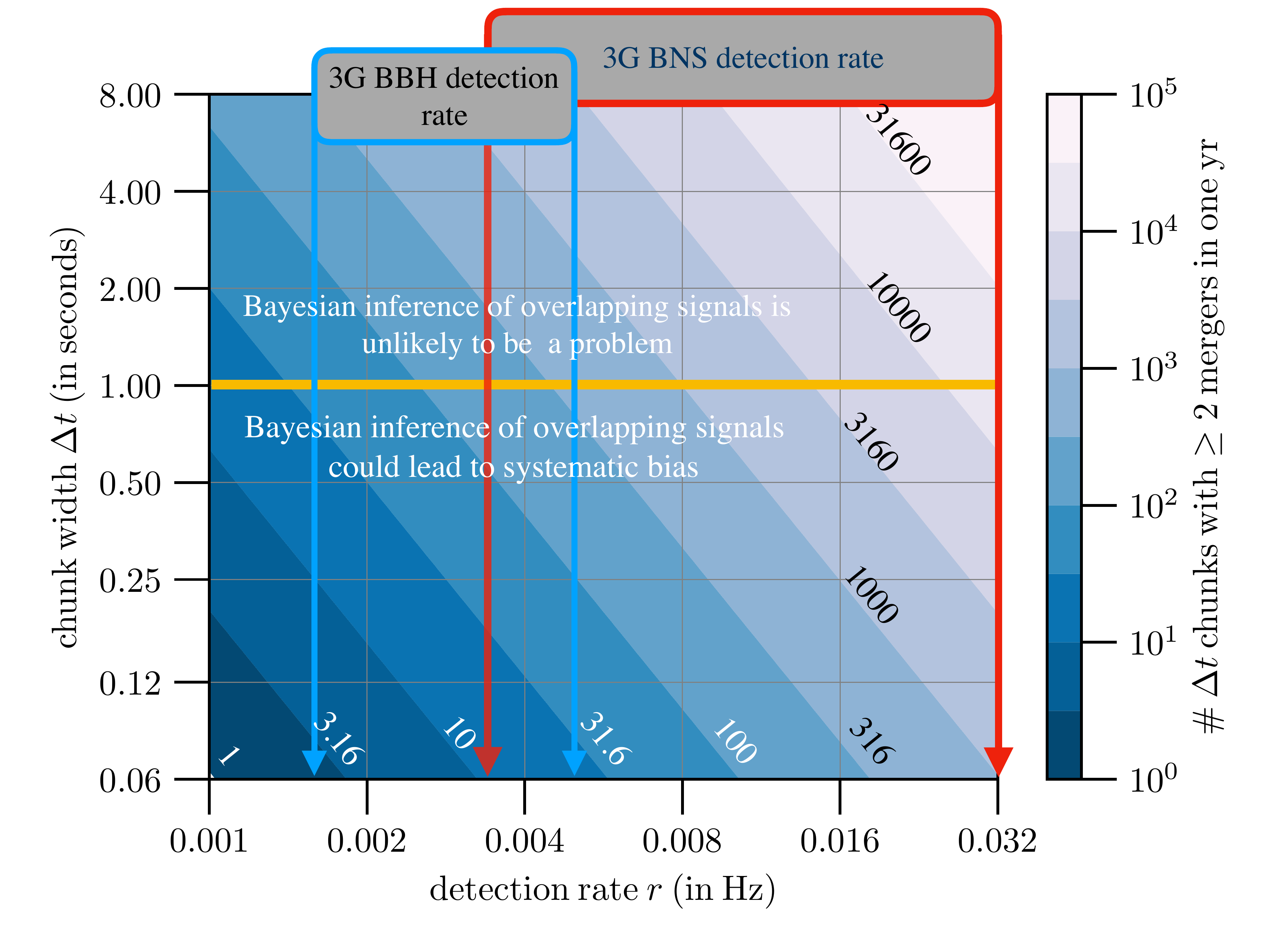}
    \caption{Contour diagram showing the number of times two or more signals have their epoch of coalescence occurring within an interval $\Delta t$ in a year's worth of data as a function of the chunk size $\Delta t$ and the Poisson rate $r.$ Also shown are the detection rate of BBH and BNS signals in 3G observatories of one ET and 2 CEs \cite{Samajdar:2021egv}. 
    As an example, if the detection rate is 8 mHz then we can expect in one year's of data 1000 one-second long chunks in which two or more mergers would occur. For a pair of signals whose coalescence times differ by an interval of $\Delta t > 1\,\rm s$ we do not expect to see any biases in their parameter estimation, although the signals themselves might overlap. Biases begin to show up for $\Delta t < 1\,\rm s$ and become severe as $\Delta t\rightarrow 0.$}
    \label{fig:overlapping signals}
\end{figure*}

The rest of the paper is organized as follows: in Sec.\,\ref{sec:overlapping chunks}, we compute the number of chunks in a year's worth of data containing more than one merger.  Section\,\ref{sec:fisher} is devoted to studying the covariance between overlapping signals using the Fisher information matrix with the emphasis on what we might expect for parameter inference in case of overlaps. Bayesian inference of overlapping signals is presented in Sec.\,\ref{sec:bayesian inference}. Our main conclusions and a brief discussion of the type of problems that should be addressed in future studies is presented in Sec.\,\ref{sec:conclusions}.

\section{Number of overlapping signals}
\label{sec:overlapping chunks}
The number of overlapping signals depends on (a) the typical duration of signals and (b) the rate at which they arrive at the detector. At the leading order, the length $\xi$ of a coalescing compact binary signal starting from a gravitational-wave frequency $f_s$ until merger is given by
\begin{equation}
\xi = \frac{5}{256}\left (G{\cal M}/c^3 \right )^{-5/3} \left (\pi f_s \right )^{-8/3}, 
\end{equation}
where $G$ is Newton's constant, $c$ is the speed of light and the chirp mass ${\cal M}$ is related to the component masses $m_1$ and $m_2$ via ${\cal M}\equiv (m_1\,m_2)^{3/5}/(m_1+m_2)^{1/5}.$ A BNS system consisting of a pair of $1.4\,M_\odot$ would last for $\xi \simeq 10^3\,\rm s$ starting from a frequency of $f_s=10\,\rm Hz$ (relevant for Advanced LIGO and Advanced Virgo), 1.8 hr for $f_s=5\,\rm Hz$ (CE) and almost 7 hr for $f_s=3\,\rm Hz$ (ET). A source of intrinsic chirp mass $\cal M$ at a cosmological redshift of $z$ would appear in the detector to have a chirp mass of $(1+z)\cal M,$ and hence lives for a shorter duration in a detector's sensitivity band.  Thus, BNSs ($1 M_\odot \le m_1, m_2 \le 3 M_\odot$) could last for tens of minutes to several hours in band while BBH signals ($3 M_\odot \le m_1, m_2 \le 50 M_\odot$) could last for tens of seconds to thousands of seconds. 

The cosmic merger rate of compact binary coalescences determined by the first two observing runs of LIGO and Virgo \cite{LIGOScientific:2018jsj, LIGOScientific:2020kqk} implies that in a network of 3G observatories the detection rate $r,$ defined as the number of signals whose matched filter signal-to-noise ratio is larger than $12$, lies in the range $r_{\rm BBH} \in [5\times 10^4, 1.5\times 10^5]$~yr$^{-1}$ for BBHs and $r_{\rm BNS} \in [10^5, 10^6]$~yr$^{-1}$ for BNSs \cite{Samajdar:2021egv,LIGOScientific:2017zlf, LIGOScientific:2016fpe}.
Thus, given that signals last for several hours, 3G data would contain several loud overlapping signals at any one time. We shall see below that for the purpose of parameter inference the relevant quantity is not how many overlapping signals there are at any one time but if two or more signals have their merger times lie within a duration $\Delta t.$ 
This is what we will set out to compute next.

\subsection{Overlapping signals of the same family}
Let $r$ denote the Poisson detection rate of a given signal family (BBH or BNS). In an interval $\Delta t$, the expected Poisson rate is $\nu = r\,\Delta t$ and the probability of observing exactly $k$ mergers during $\Delta t$ is given by
\begin{equation}
    P_k(\nu) = \frac{\nu^k\,e^{-\nu}}{k!}.
\end{equation}
Thus, the probability of observing two or more mergers during $\Delta t$ is 
\begin{equation}
    P_{k\ge 2} = \sum_{k=2}^\infty P_k(\nu) =
                 \sum_{k=2}^\infty  \frac{\nu^k\,e^{-\nu}}{k!}
                 = 1 - e^{-\nu}(1+\nu).
\end{equation}
We have made use of the fact that the Poisson distribution is normalized, namely $\sum_{k=0}^\infty P_k(\nu)=1.$ To compute the number of chunks $N_{k\ge 2}$ in which two or more mergers will be observed we must multiply the probability $P_{k\ge 2}$ by the number of chunks $n_{\Delta t}=T/\Delta t$ in an observational period $T$:
\begin{equation}
    N_{k\ge 2} \equiv P_{k\ge 2} n_{\Delta t} = \left [ 1 - e^{-\nu}(1+\nu) \right ] \frac{T}{\Delta t}.
\end{equation}
Substituting $\Delta t = \nu/r$ and noting that $N_T \equiv r\,T$ is the total number of signals detected during the period $T,$ we get
\begin{equation}
    N_{ k\ge 2} 
    = \left [ 1 - e^{-\nu}(1+\nu) \right ] \frac{N_T}{\nu}.
\end{equation}
It is easy to see that in the limit $\Delta t\rightarrow 0$ (equivalently, $\nu \rightarrow 0),$  $N_{ k\ge 2} \simeq \nu N_T/2.$ The factor of 1/2 assures that the number of instances when two or more signals are found in a chunk is never greater than half of the total number of observed signals but it is also weighed down by the Poisson rate $\nu.$  In the other limit, when $\Delta t\rightarrow T$ (and $\nu\gg 1$), $N_{k\ge 2} \simeq 1$ but less than 1. 

Figure \ref{fig:overlapping signals} plots the number of chunks $N_{k\ge 2}$ in which we can expect to find two or more mergers in a year's worth of data (i.e., using $T=1\,\rm yr$ and $\nu = r\, \Delta t$). Also indicated in the plot are the detection rate of BBH (BNS) which is expected to be in the range $r_{\rm BBH} \in [1.6, 4.8]\times 10^{-3}\,\rm s^{-1}$ ($r_{\rm BNS} \in [3.5, 35]\times 10^{-3}\,\rm s^{-1}$, respectively) \cite{Samajdar:2021egv}  
in a 3G detector network comprising of one ET and two CEs (one in north America and the other in Australia). As we shall see in Sec.\,\ref{sec:fisher}, parameter inference should not be a problem if the difference in coalescence times of a pair of signals is larger than $\sim 1\,\rm s$; this is indicated in Fig. \ref{fig:overlapping signals} by the horizontal line drawn at $\Delta t=1\,\rm s.$ Thus, in Sec. \ref{sec:bayesian inference} we will focus on Bayesian inference of signals whose merger times differ by about one second. We see that at the higher end of the BNS rate, we expect $\sim 15,000$ one-second long chunks with two or more mergers while at the lower end of the BNS rate this number is $\sim 200.$  Likewise, $\sim 300$ chunks will contain two or more BBH mergers at the higher end of the BBH detection rate while this number is $\sim 40$ at the lower end of the BBH rate. Although the vast majority of events will have their merger times larger than 1 s from their nearest neighbor, the number of events with their merger times within a second is quite large.

The detection rate of BBH signals in the current detector network of LIGO, Virgo and KAGRA at their design sensitivity is at best $r \sim 2.3\times 10^{-5}$~s$^{-1}$ (or 730~yr$^{-1}$) 
\cite{LIGOScientific:2020kqk}. Thus, the probability of observing multiple mergers in a chuck of size 1 s or less is negligibly small in the Advanced detector era. This will also be the case in the A+ era \cite{KAGRA:2013rdx} where the detection rates are expected to be $3$ times larger. 

\subsection{Overlapping signals from two different families}
If the detection rate of signal families A and B are  $r_A$ and $r_B$, then probability that \emph{one} or more mergers of each of these signal families would occur during an interval $\Delta t$ is 
\begin{equation}
    P_{A, k\ge 1} = 1-e^{-\Delta t\, r_A},\quad   P_{B, k\ge 1} = 1-e^{-\Delta t\, r_B}. 
\end{equation}
Thus, the probability $P_{AB}$ that an interval $\Delta t$ contains one or more from each of the two signal families is simply the product $P_{AB} = P_{A, k\ge 1}\, P_{B, k\ge 1.}$ If the rates are small, this reduces to $P_{AB} = (\Delta t)^2\,r_A\,r_B$ and the number of such chunks over a period $T$ is $N_{AB}= (\Delta t)^2 r_A\,r_B\,T = N_A\,N_B/n_{\Delta t},$ where $N_A$ and $N_B$ are the total number of mergers during the period $T$ of families $A$ and $B,$ respectively, and $n_{\Delta t}=T/\Delta t$ is the number of chunks of width $\Delta t$ during $T.$ Using the range of BNS and BBH rates quoted before, we find that $N_{AB}$ would lie in the range 170--5100 for $T=1\,\rm yr$ and $\Delta t =1 \,\rm s.$ 

From the foregoing discussions it is clear that a small but  significant fraction of signals would have their coalescence time within an interval of 1 s. As we shall see in the next Section, due to their long duration, overlapping BNS signals are far less correlated with each other than overlapping BBH signals. For the same reason, a pair of overlapping BNS and BBH signals are poorly correlated. Hence, in the Bayesian inference problem (Sec.~\ref{sec:bayesian inference}) we will only consider overlapping BBH signals.

\begin{figure*}
\begin{center}
\includegraphics[width=0.9\textwidth]{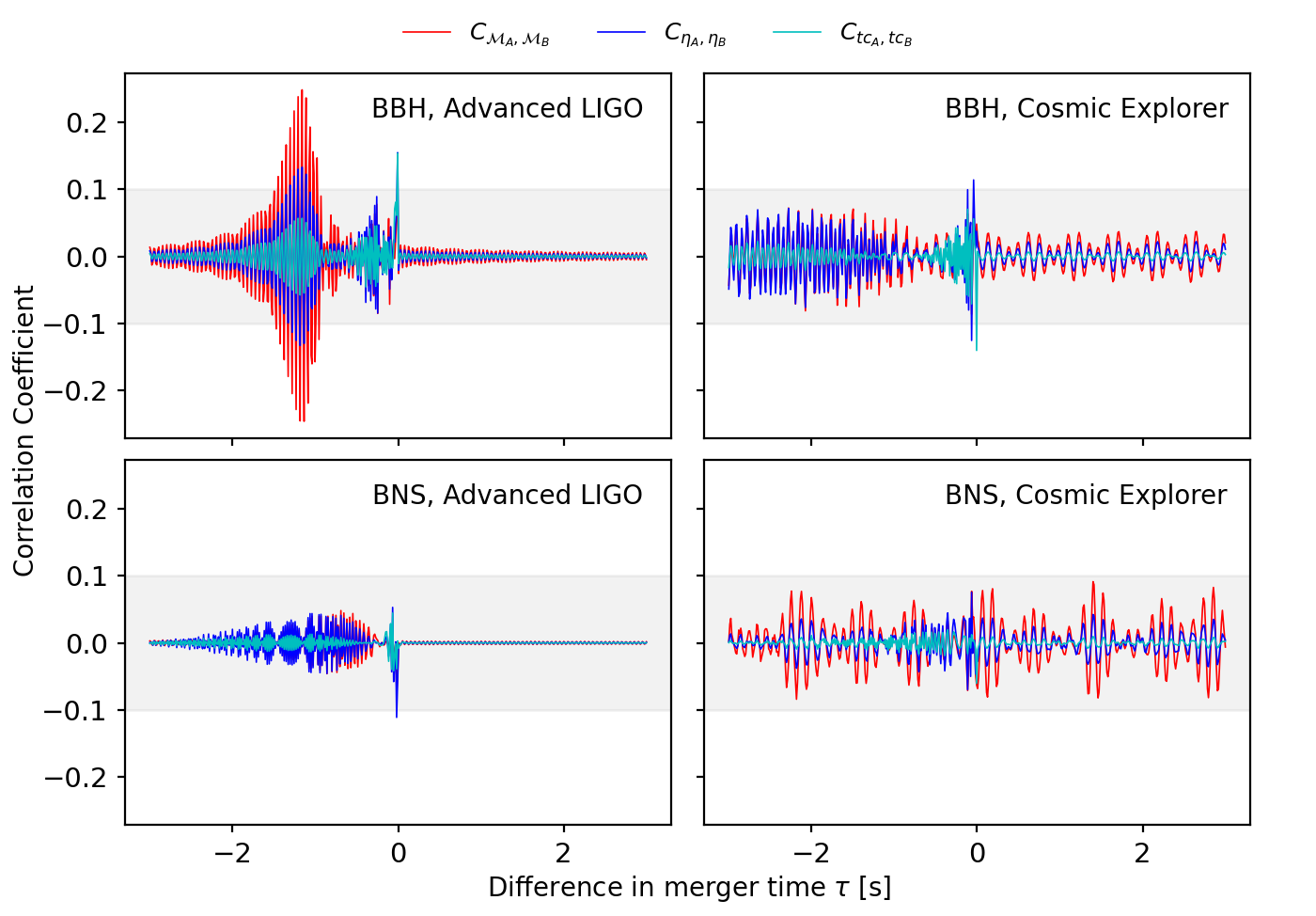}
\end{center}
\caption{Plot shows the correlation coefficients, i.e., normalized covariances as defined by 
Eq.\,(\ref{eq:cc}) 
between the parameters of the two overlapping signals as a function of the difference in merger times $\tau=t_C^B-t_C^A.$ The left panel is for Advanced LIGO and right for Cosmic Explorer. Top row is for BBHs and bottom row BNSs. We assume the parameter inference of overlapping signals to be negligibly affected when (the absolute value of) the correlation coefficients are less than 10\% (grey shaded regions).
}
\label{Fig:fisher_correlation_BBH}
\end{figure*}

\section{Covariance among overlapping signals}
\label{sec:fisher}
If two signals are well separated then the covariance between their parameters is zero and we do not expect one signal to affect the parameter inference of the other. As we bring the two signals closer together in time, at some point the presence of one of the signals will begin to bias the estimation of parameters of the other.  In this Section we estimate the covariance between the parameters of a pair of overlapping signals using the Fisher matrix formalism. Although Fisher matrix is valid in the limit of large signal-to-noise ratios, any inferences we can draw from the correlation will guide us in choosing the parameter space of compact binaries where systematic biases could be large.

To this end, we assume that the data contains a pair of signals $s_A$ and $s_B$ buried in stationary, Gaussian noise $n.$ The detector output is a sum of the overlapping signals buried in detector noise:
\begin{equation}
    x(t) = n(t) + s_A(t, \lambda^{(A)}_\alpha) + s_B(t, \lambda^{(B)}_\alpha). 
\end{equation}
where $\lambda^{(A)}_\alpha,$ $\lambda^{(B)}_\alpha,$ for  $\alpha=1,\ldots, p,$ are the set of parameters corresponding to signals $s_A$ and $s_B,$ respectively. Note that since both $s_A$ and $s_B$ are assumed to belong to the same signal family they are specified by the same number of parameters. Furthermore, we shall only consider a single detector for this exercise. The relevant parameters for a binary with non-spinning companions are  the chirp mass $\cal M$, symmetric mass ratio $\eta$, the epoch $t_C$ when the signal amplitude reaches its peak and the phase $\phi_C$ of the signal at that epoch and so:
$\lambda^{(A)}_\alpha = ({\cal M}^{(A)}, \eta^{(A)}, t_C^{(A)}, \phi_C^{(A)})$ and similarly for signal $s_B.$  We assume the {\sc IMRPhenomPv2} waveform model.

For the computation of the covariance matrix it is more convenient to consider that the data contains only one signal, i.e., the sum of the two signals $s=s_A+s_B$, and it is characterized by a double number of parameters: $\theta_a=\lambda^{(A)}_a$ for $a=1,\ldots, p$ and $\theta_a=\lambda^{(B)}_{a-p}$ for $a=p+1,\ldots, 2p.$  For a noise background that is stationary and Gaussian the covariance matrix $C,$ which is inverse of the Fisher matrix $\Gamma,$ is given by:
\begin{equation}
    C_{ab} = \Gamma^{-1}_{ab},\quad
    \Gamma_{ab} =  \left < \frac{\partial s}{\partial \theta_a}, \frac{\partial s}{\partial \theta_b} \right >.  
\end{equation}
Here the scalar product of two waveforms (or any pair of functions of time for that matter) $h$ and $g$ is defined as
\begin{equation}
    \left <h, g \right > \equiv 4 \Re \int_{f_{\rm low}}^{f_{\rm high}} \frac{\tilde h(f)\,\tilde g^*(f)}{S_h(f)}\,{\rm d}f,
\end{equation} 
where $\Re$ stands for the real part of the integral, $\tilde h$ and $\tilde g$ are the Fourier transforms of the signals $h$ and $g,$ respectively, $g^*$ denotes the complex conjugate of $g$ and $S_h(f)$ is the one-sided noise spectral density of the detector. In our study we will use either the noise spectral density of Advanced LIGO \cite{LIGOScientific:2014pky} or that of the Cosmic Explorer \cite{Reitze:2019dyk}.
The lower frequency cutoff $f_{\rm low}$ is chosen to be 20 Hz for Advanced LIGO and 5 Hz for Cosmic Explorer. For BNSs, the upper frequency cutoff $f_{\rm high}$ is assumed to be the larger of the inner-most stable circular orbit frequency of the two overlapping signals, i.e.,  $f_{\rm high} = {\rm max}[(6^{3/2}\pi M_1)^{-1}, (6^{3/2}\pi M_2)^{-1}]$, where $M_1$ and $M_2$ are the total mass of the two overlapping signals. For BBHs, the upper frequency cutoff is chosen to be the Nyquist frequency of 1024 Hz. 

The Fisher matrix contains interference terms of the following type: 
\begin{equation}
    \Gamma_{\alpha,\,\beta+p} = \left < \frac{\partial s_A}{\partial \lambda^{(A)}_\alpha}, \frac{\partial s_B}{\partial \lambda^{(B)}_\beta} \right >.  
\end{equation}
Covariances are of primary interest in this Section as they can tell us the degree to which the presence of one signal affects the parameter inference of the other. In order to measure the extent of covariance we consider two sets of overlapping signals (masses are specified in the detector frame):
\begin{enumerate}
    \item overlapping BBHs with masses: 
    \begin{eqnarray}
(m_1^{{(A)}},\,  m_2^{{(A)}}) & = & (21\,M_\odot,\, 15\,M_\odot) \\
(m_1^{{(B)}},\,  m_2^{{(B)}}) & = & (33\,M_\odot,\, 29\,M_\odot). 
\label{eq:BBH masses}
    \end{eqnarray}
    
    \item overlapping BNSs with companion masses:
\begin{eqnarray}
(m_1^{(A)},\, m_2^{(A)})& = &(1.45\, M_{\odot}, 1.35\, M_{\odot})\\  
(m_1^{(B)},\, m_2^{(B)})& = &(1.50\, M_{\odot}, 1.40\, M_{\odot}).
\label{eq:BNS masses}
\end{eqnarray}
\end{enumerate}
Furthermore, in all cases we choose
\begin{eqnarray}
(t_C^{(A)}, \phi_C^{(A)}) = (0,\,  0),\quad 
(t_C^{(B)}, \phi_C^{(B)}) = (\tau,\, 0), \label{eq:tau_def}
\label{eq:merger times}
\end{eqnarray}
and vary $\tau$ over the range {\red $[-3,\, 3]\,\rm s.$}

The covariances between the chirpmass, symmetric mass ratio and epoch of coalescence are plotted in Fig.\,\ref{Fig:fisher_correlation_BBH} as a function of the parameter $\tau$ for overlapping BBHs (top panels) and BNSs (bottom panels) for noise spectral densities of Advanced LIGO (left panels) and Cosmic Explorer (right panels). Other cross-covariances are negligibly small and not shown. What we plot are the normalized covariances, i.e., a combination of the correlation coefficients defined as:
\begin{equation}
    \sigma_{ab} \equiv \frac{C_{ab}}{\sqrt{C_{aa}C_{bb}}},\quad a\ne b.
    \label{eq:cc}
\end{equation}
This quantity is strictly bounded between $-1$ and $+1.$ 
A correlation coefficient of $+1$ implies that the parameters are perfectly correlated, $-1$ implies they are perfectly anti-correlated, and a value of 0 would imply they are uncorrelated. We will take $\sigma_{ab}\sim 0.1$ (grey shaded region in the plot) to be small enough to indicate that the presence of the second signal does not significantly bias parameter inference of the other signal. This threshold is inevitably arbitrary, as a thorough analysis of the connection between the values of the correlation coefficients and the presence of biases in parameter inference is beyond the scope of this paper. However, as we show in Sec. \ref{sec:results}, the regions of the parameter space where biases in PE arise are compatible with the ones for which $\sigma_{ab}\gtrsim 0.1$.  

The correlations displayed in Fig. \ref{Fig:fisher_correlation_BBH} show a range of different behaviours. In all cases, they have a peak for $|\tau| \lesssim 0.5\,\rm s$. This is expected, as the interaction between the signals is enhanced when the two signals coalesce very close to each other. For $|\tau| > 0.5\,\rm s$, all the different configurations stay always below the threshold $\sigma_{ab}= 0.1$, with the significant exception of BBH in Advanced LIGO detectors. In this latter case, correlations remain very high in the range $-1.5\,\mathrm{s}<\tau<0\,\mathrm{s}$, and become small only for $\tau\lesssim -2\,\mathrm{s}$. The fact that correlations are not symmetric in $\tau$ can be easily explained by the different form of the two signals considered (see also Fig. \ref{fig:waveforms}).

Finally, we note that in the case of BNS, the correlation remains always below the threshold both in Advanced LIGO and Cosmic Explorer, except when $\tau\simeq 0$. This implies that parameter inference of overlapping BNS signals is likely to be less severe than that of overlapping BBH signals. We will, therefore, consider only the latter class of signals in the remainder of this paper, leaving the parameter estimation of overlapping BNS signals for future work. 

The analysis presented in this section is limited by the fact that we have explored only for a few particular sets of source parameters. Therefore, we cannot conclude that parameter estimation will never be a problem in the case of overlapping BNSs. Indeed, very similar values of the chirp masses (as well as other relevant parameters) will likely increase the correlation between the two signals, especially in the proximity of $\tau=0\,\mathrm{s}$.

In addition, we note that further work is necessary to assess the validity of the correlation threshold we have considered here, especially in light of the fact that sinusoidal features with amplitudes $\sigma_{ab}\approx 0.05-0.1$ are present in the case of the Cosmic Explorer detector, even for large values of $|\tau|$. Despite the fact that these correlations are very low, their effects on the results of parameter inference need to be evaluated quantitatively.

\section{Bayesian inference of overlapping signals}\label{sec:bayesian inference}

In this Section, we support the results we have derived using the Fisher information matrix formalism (Sec. \ref{sec:fisher}) with a full Bayesian inference procedure. With this \textit{parameter estimation} (PE) process, we are able to fully explore the posterior distribution of the parameters that generated the signals. This is important, because it allows us to confirm the presence (expected from the Fisher study) of distinct maxima in the posterior, one for each signal coalescing within the time chunk considered. Moreover, thanks to this numerical approach, we can explore more carefully the region where biases are expected assessing their significance and gauging the conditions for which they seem to happen.

Within the Bayesian framework, given a set of parameters $\lambda$ describing a compact binary coalescence (CBC) waveform $h(\lambda, t)$, we can write the posterior distribution for $\lambda$ as:
\begin{equation}
    P(\lambda\,|x,h) = \frac{\pi(\lambda)\,\mathcal{L}(x\,|\lambda, h)}{\mathcal{Z}(x)},
\end{equation}
where $x$ is the detector output. This posterior can be explored by using a sampling algorithm (e.g., MCMC, nested sampling). 
As in Sec. \ref{sec:fisher}, assuming that the data $x$ contains two overlapping signals $s_A$ (\textit{signal A}) and $s_B$ (\textit{signal B}), then it can be written as:
\begin{equation}
    x = n + s_A + s_B,\label{eq:stream}
\end{equation}
where $n$ is the noise of the interferometer. Note that, in principle, to perform a Bayesian analysis of two or more overlapping signals we should broaden the parameter space, e.g., $\theta=\{\lambda^A,\lambda^B\},$ in order to account for the presence of multiple overlapping signals. However, since running a sampling algorithm requires significant amount of computational resources, in most cases this is not required. In fact, as argued in Sec.\,\ref{sec:fisher}, if the signals' coalescence times are wide apart we do not expect the presence of one signals to influence posterior distribution of parameters of the other. For this reason, in what follows we consider the parameter space of a single CBC signal. We will return to this point later on when discussing possible biases arising because of this choice. 

\subsection{Choice of signal families} \label{sec:parameters_run}

\begin{figure*}
\includegraphics[width=1.0\textwidth]{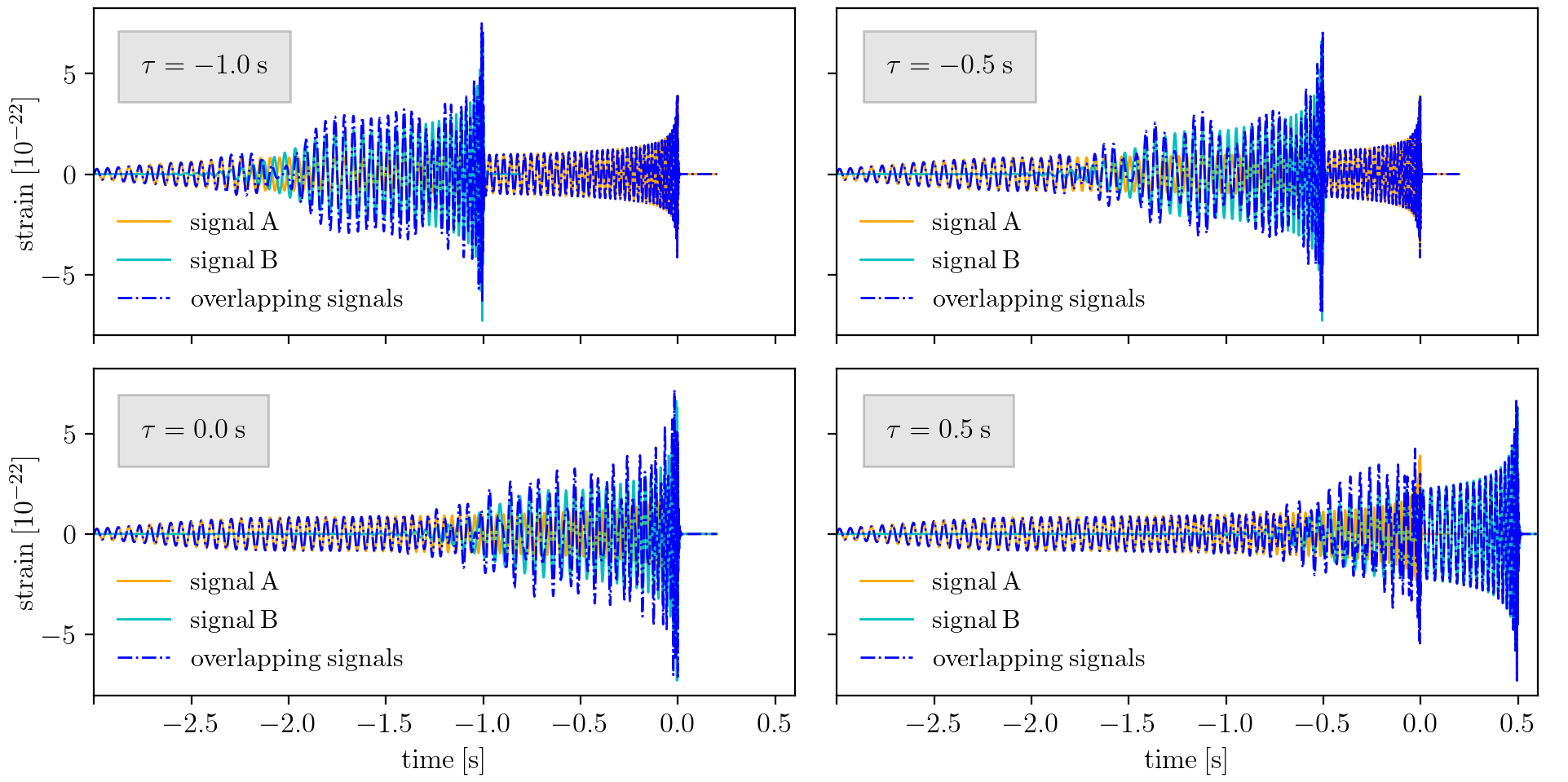}
\caption{Signals in the time domain, for four different values of the time shift $\tau$. \textit{signal A}, \textit{signal B}, and the resulting overlapping signal are plotted. The waveforms are generated using the approximant {\sc IMRPhenomPv2}. The two luminosity distances are fixed to $d_L^{(A)} = 1\,\mathrm{Gpc},\,d_L^{(B)} = 1\,\mathrm{Gpc}$. Note that, if we neglect the effects of cosmological redshift, then changing these distances just results on a scaling of the signals' amplitude.}
\label{fig:waveforms}
\end{figure*}

As already mentioned, in this analysis we focus only on BBH signals. This choice is motivated by the fact that: (a) covariances among overlapping BNS signals are smaller than the BBH ones (Sec.\,\ref{sec:fisher}), and, therefore, biases in the BNS case are expected to be less important; (b) BNS signals last for several hours in 3G detectors and tens of minutes in Advanced LIGO and Virgo, implying that Bayesian inference takes a formidable amount of computational resources (although new algorithms are already showing the promise of greatly reducing the computational requirement \cite{Zackay:2018qdy, Cornish:2010kf, Finstad:2020sok}).

Furthermore, we also restrict our analysis using Advanced LIGO sensitivity. As argued before, LIGO is not affected by the problem of overlapping signals, because the rate and the duration of the signals are far too small to create any overlap. Nonetheless, in this work we are not really interested in reproducing a realistic set of overlapping data; instead, we want to focus on the parameter estimation process. To do so, there is no substantial advantage in using 3G mock data: we expect that our conclusions will be valid even if they are based on the analysis of Advanced LIGO mock data.

The parameters of the overlapping BBH signals used in Bayesian inferences is the same as what we used in Sec.\,\ref{sec:fisher}: nonspinning BBHs with masses as given in Eq.~(\ref{eq:BBH masses}) and coalescence times and phases as given in Eq.~(\ref{eq:merger times}).  We ignore the position of the sources in the sky and their orientation relative to the detectors (setting all angles to zero). 
We do, however, include in our analysis the luminosity distance $d_L$ of the source. The parameter space we use in our analysis is thus:
\begin{equation*}
    \lambda = \{m_1, m_2, \phi_C, t_C, d_L\} 
\end{equation*}
Note that our choice of sky position is the worst case scenario, because we are considering the two sources to have the same exact location in the celestial sphere. In reality, if overlapping signals arrive from different directions in the sky, they will have different phase coherence amongst a network of detectors and thus easier to discriminate. Thus, since our choice of sky position is the worst case scenario, the parameter estimation problem can only be better when sky position and orientation are taken into account.

To explore different configurations of the parameters, we vary the time shift $\tau$ - defined in Eq.\,(\ref{eq:tau_def}) as the epoch coalescence of \textit{signal B} - in the range $\tau \in \{-2.0\,\mathrm{s}, 0.5\,\mathrm{s}\}$.
Along with the time shift, we also vary the two luminosity distances of the sources $d_L^{(A)}$ and $d_L^{(B)}$, and their phases $\phi_C^{(A)}$ and $\phi_C^{(B)}$. In the first set of runs, we fix $\phi_C^{(A)} = \phi_C^{(B)} = 0$ and vary the two distances. We keep the distance of one of the sources fixed to $1\,\mathrm{Gpc}$ and set the other at either 500 Mpc, 1 Gpc, or 2 Gpc. In the second set of runs, we vary the phase of signal B ($\phi_C^{(B)}\in\{0,\pi/3, 2\pi/3\}$), keeping $\phi_C^{(A)}$ fixed to zero and the two luminosity distances to $d_L^{(A)} = d_L^{(B)} = 1\,\mathrm{Gpc}$.

The resulting variations in the parameter sets are:
\begin{eqnarray}
    \tau &= & \{-2.0\,\mathrm{s}, -1.5\,\mathrm{s}, -1.0\,\mathrm{s}, -0.5\,\mathrm{s}, 0.0\,\mathrm{s}, 0.5\,\mathrm{s}\}\\
    d_L^{(B)},\, d_L^{(A)} & = & \{500\,\mathrm{Mpc}, 1\,\mathrm{Gpc}, 2\,\mathrm{Gpc}\}\\
    \phi_C^{(B)}& = & \{0, \pi/3, 2\pi/3\}
\end{eqnarray}
With these choices, there are 48 different possible configurations, each of which is analyzed for Bayesian parameter inference.

In the inference problem we use a signal model that accurately represents the BBH waveforms. As in Sec. \ref{sec:fisher}, we use the {\sc IMRPhenomPv2} approximant to create waveforms in the frequency domain, fixing the low frequency cutoff to be $20\,\mathrm{Hz}$, which is consistent with the minimum frequency used in the LIGO/Virgo PE. In Fig.\,\ref{fig:waveforms}, we plot the two waveforms in the time domain, for the different configurations of the time shift $\tau$. 
The resulting overlapping waveform is plotted as well. In Table \ref{tab:SNR}, we compute the expected matched filter SNR for the different possible configurations of the parameters, focusing on the distances, since neither the coalescence time nor the phase affect the SNR value.
\begin{table}[h!]
    \centering
    \begin{tabular}{c|c|c|c}
\hline\hline       
 SNR & $d_L = 0.5\,\mathrm{Gpc}$ & $d_L = 1\,\mathrm{Gpc}$ & $d_L = 2\,\mathrm{Gpc}$\\
    \hline
    signal A  & 54.2 &  27.1  &  13.5\\
    signal B  & 82.8 &  41.5  &  20.7 \\
\hline\hline       
       \end{tabular}
    \caption{SNRs for the two signals we have chosen to focus on in our analysis (considering the two LIGO interferometers network), created with different values of the luminosity distances $d_L$. Note that applying a time shift to the signals do not change the value of the SNR.}
    \label{tab:SNR}
\end{table}

\subsection{Setting up Bayesian inference runs}
Having created the mock data with overlapping signals we next focus on parameter inference.  Our analysis uses two LIGO interferometers, but our conclusions are not significantly affected by this choice: considering a different detector network would simply result in different SNRs for the signals as we are not focusing on the sky position of the source. Although this could in principle change the heights of the peaks in the posterior distribution, we do expect it to influence their relative ratios significantly, and hence the PE process we consider is expected to hold for any network. 

The data set consists of 4 s of mock data from the two LIGO interferometers. $4\,\mathrm{s}$ is large enough to span the full length of the longer signal. We do not add any noise to the data -- i.e., we set $n=0$ in Eq. (\ref{eq:stream}) --, as we want to highlight the presence of biases created by the overlap between the signals, and these biases could be covered by the statistical uncertainty created by the presence of noise.  

We use the \code{Bilby} package to perform Bayesian parameter inference of the two signals, running the \code{dynesty} sampler \cite{Speagle:2019ivv}. \code{Dynesty} is a dynamic, nested sampling algorithm \cite{Skilling:2006gxv, Higson_2018}, which is well suited for our purposes because it quickly achieves convergence, but at the same time it is able to handle non-trivial, multi-modal distributions better than MCMC-based algorithms \cite{Speagle:2019ivv}. We allow the sampler to explore the likelihood surface with respect to all the parameters except $\phi_C,$ over which the likelihood is analytically marginalized, and $d_L,$ over which the likelihood is numerically maximized. Marginalization over $\phi_C$ and $d_L$ correctly accounts for the effects of the parameters $\phi_C$ and $d_L$ on the resulting 3-d posterior \cite{Veitch:2014wba, Singer:2015ema}.  

\subsection{Bayesian priors} \label{sec:priors}
At the beginning of the analysis, we have to set the priors on the various parameters. We consider a uniform prior on the phase $\phi_C$, with periodic boundary conditions, a power-law prior on the luminosity distance, $p(d_L) \propto d_L^\alpha$ with $\alpha=2,$ and a uniform prior on the two masses $m_1$ and $m_2$ over the range $[10\,\msun,\, 50\,\msun]$. As for the coalescence time, selecting the best possible prior turns out to be a game-changing strategy. In fact, running a simulation with a wide prior on the time $t_C$ that spans the merger times of the two overlapping signals leads to significant problems: while one of the two signals is always recovered correctly, the other is completely ignored by the sampling algorithm. A wide prior on $t_C$, therefore, would only allow us to infer the parameters of the louder signal, without access to the weaker one. 

\begin{figure*}
\includegraphics[width=1.0\textwidth]{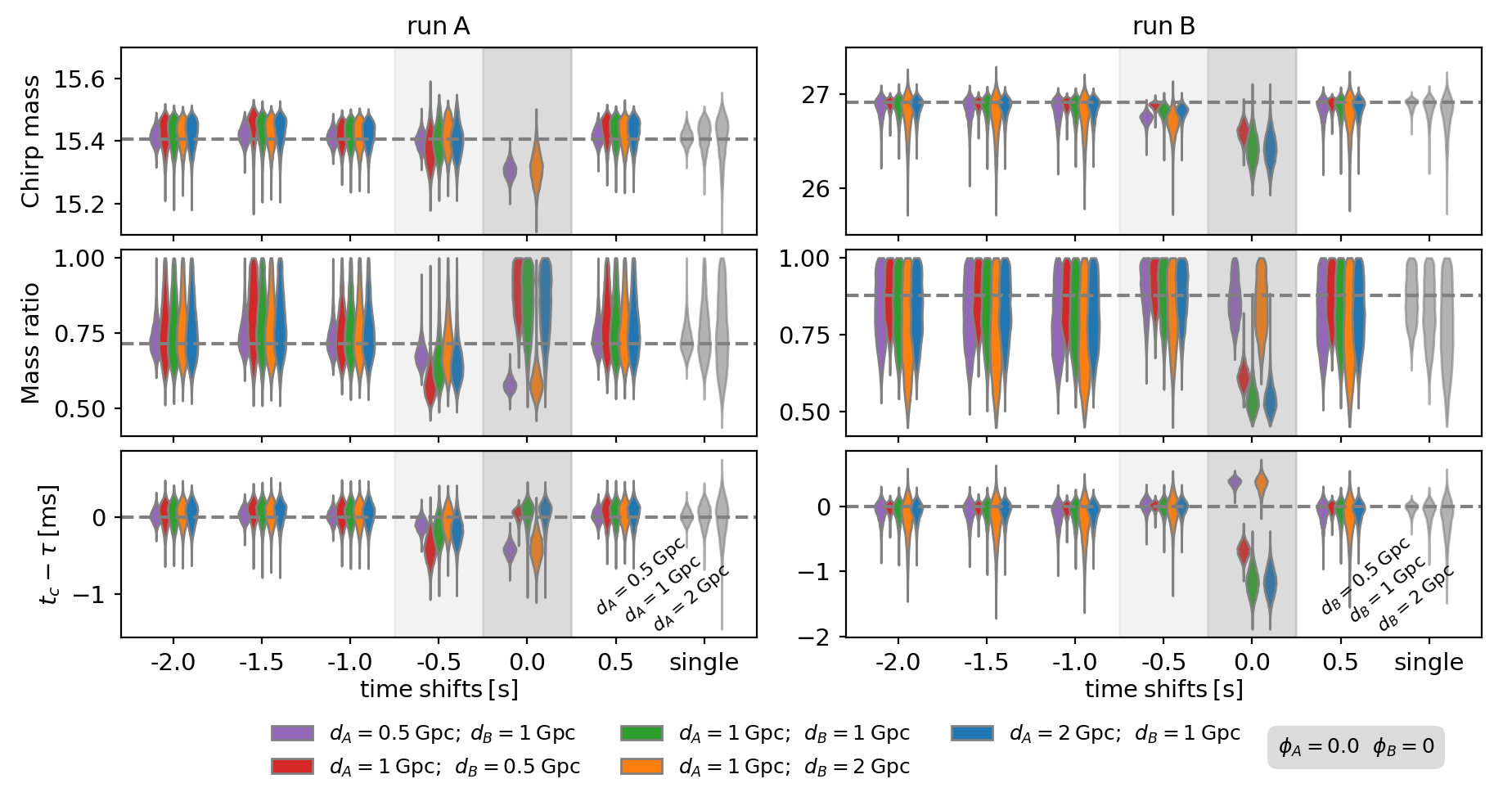}
\vspace{10pt}

\includegraphics[width=1.0\textwidth]{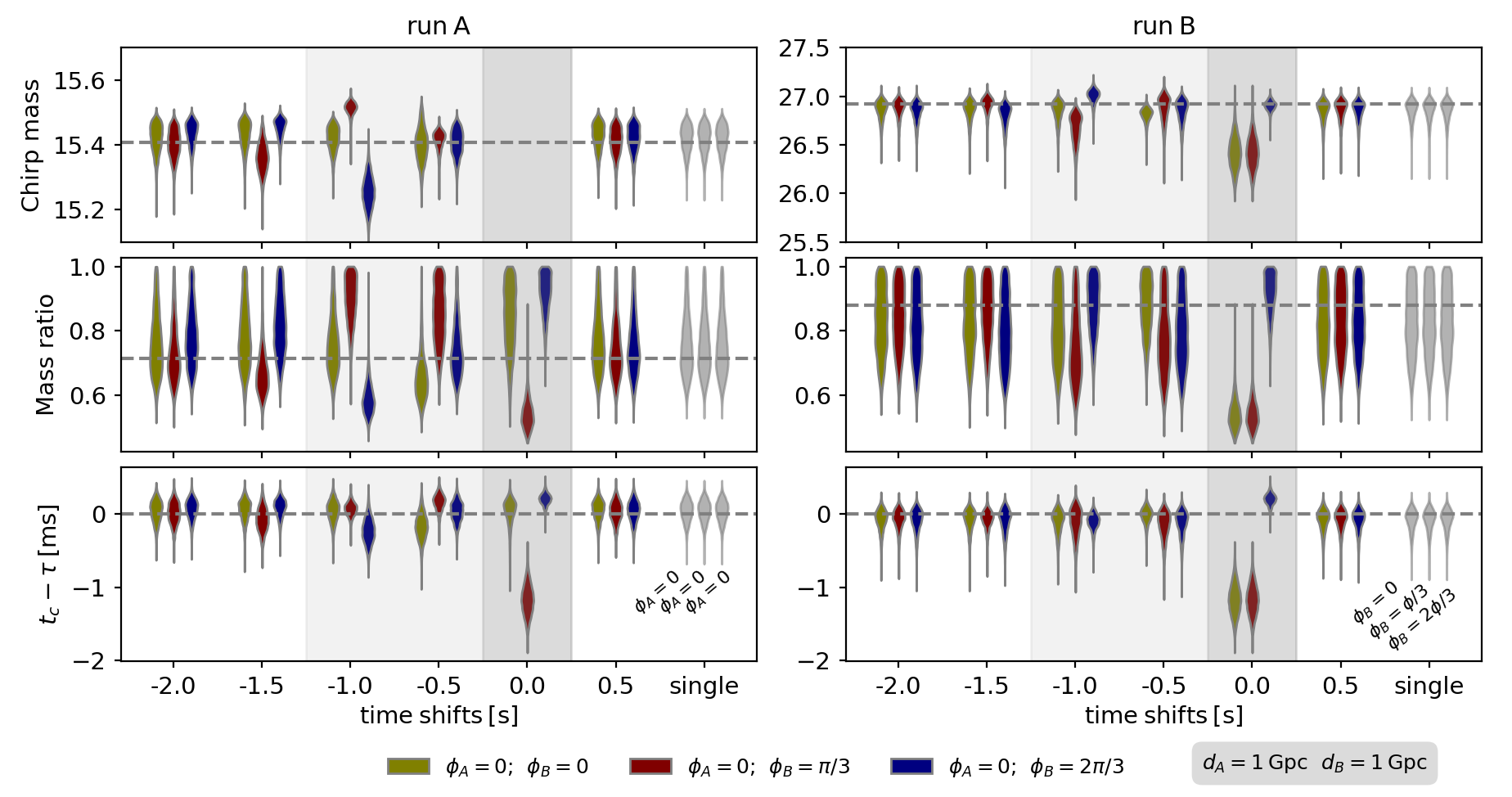}
\caption{Summary of the results for the set of 48 runs, each one with a different configuration of the parameters $\tau$, $d_L^\mathrm{(A)}$ and $d_L^\mathrm{(B)}$ (top panels), $\phi_C^{(A)}$ and $\phi_C^{(B)}$ (bottom); for details about the parameters choice, see Sec. \ref{sec:parameters_run}. \textit{A runs} are shown on the left panels, and \textit{B runs} are on the right panel. Posterior distributions for the chirp mass $\mathcal{M}$, mass ratio $q$, and coalescence time $t_C -\tau$ are shown in the form of violin plots. Along with the results for overlapping signals, posteriors for the ``\textit{single signal}" case (i.e., only one signal is present in the data) are shown in the rightmost side of each panel in grey. The true values of the masses and times for \textit{signal A} and \textit{signal B} are highlighted with dashed horizontal lines. Note that the distributions in the plots referring to the same time shift $\tau$ are slightly shifted with respect to their exact value of $\tau$ so that they do not overlap with each other. The $\tau=0.0\, \mathrm{s}$ runs are highlighted with a dark grey shadowed band; other regions where non-negligible biases are present (see discussion in Sec. \ref{sec:results}) are highlighted in the same way with a lighter shade of grey. Note that in the $\tau=0.0\,\mathrm{s}$ case, part of the recovered values for the chirp masses are out of the range and thus not shown. 
}
\label{fig:res_summary}
\end{figure*}

However, as already pointed out, previous work suggests that signals can always be detected, even if they are overlapping, and their merger time correctly identified \cite{Regimbau:2012ir, Meacher:2015rex}. Although these studies dealt only with BNS signals, we do expect that similar conclusions hold also in the case of BBH. This is because (as we show in Sec. \ref{sec:results}, Fig. \ref{fig:res_summary}) biases on the values of $t_c$ recovered from our PE analysis are minor (at the ms level) and the presence of the overlap does not seem to hamper the time recovery of the signals. However, future efforts will need to back up this assumption and confirm that BBH overlapping signals can be correctly recovered in time domain. From current pipelines, we know that the detection of a signal allows us to know its epoch of coalescence with very low uncertainty (at the order of $10\,\mathrm{ms}$). We then assume to know the time of coalescence of the two overlapping signals with a good degree of accuracy, and constrain our parameter space choosing a prior on the coalescence time which is centered on the (fiducial) true value of the time $t_C,$ with a width of $100\,\mathrm{ms}$. In this way, for each of the signals we can isolate the region of the parameter space where we expect to find the true values of the injection parameters. This choice allows us to recover the correct parameters for both \textit{signal A} and \textit{signal B}.

Therefore, for each of the 48 injections, we run the Bayesian inference procedure two times: the first one (we refer to it as \textit{run A}) aims to recover the true values of the parameters of \textit{signal A}; to this end, since $t_C^\mathrm{(A)}=0.0\,\mathrm{s}$, we set the prior on the coalescence time centered around zero. \textit{Run B}, on the other hand, focuses on the \textit{signal B} peak in the parameter space; thus, the prior is chosen to be centered in $t_C=\tau$.

\subsection{Results}
\label{sec:results}

\begin{figure*}
\begin{center}
\includegraphics[width=0.48\textwidth]{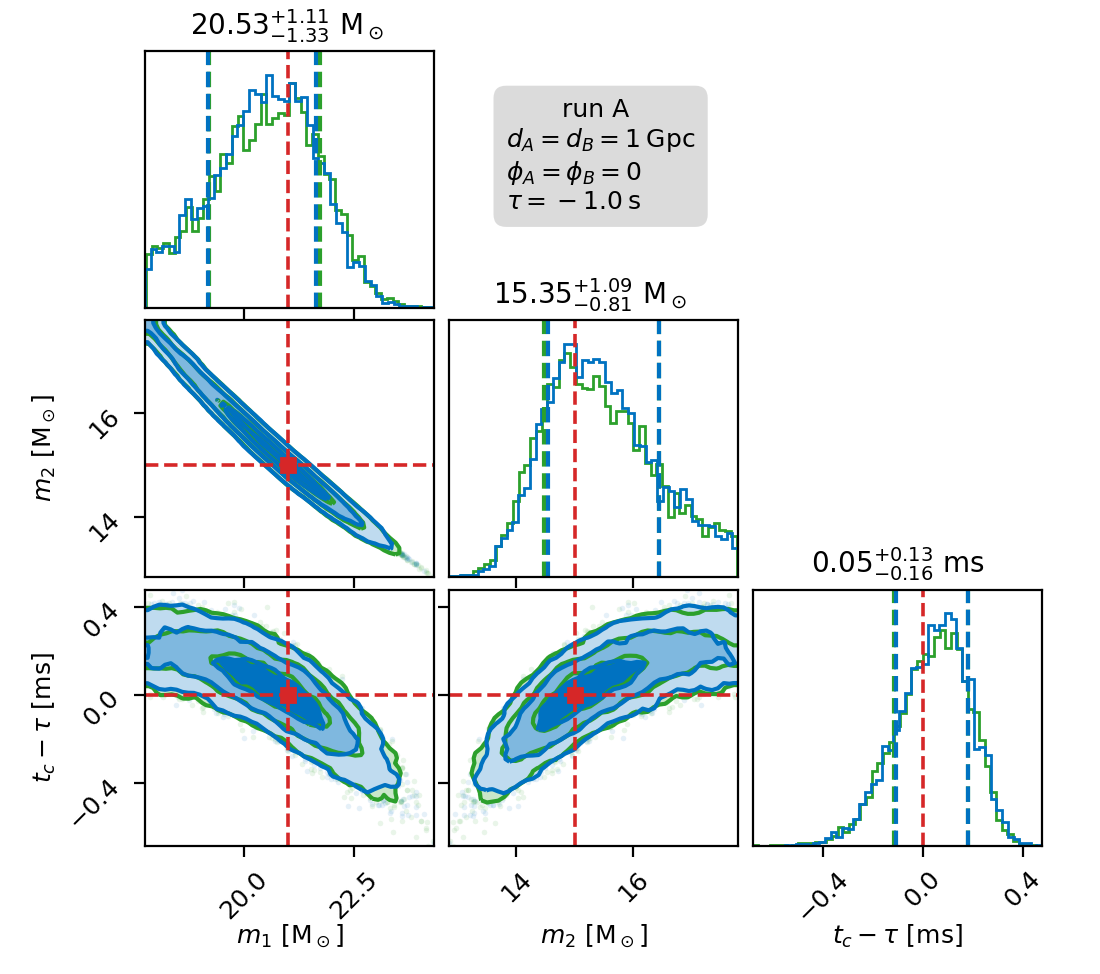}
\includegraphics[width=0.48\textwidth]{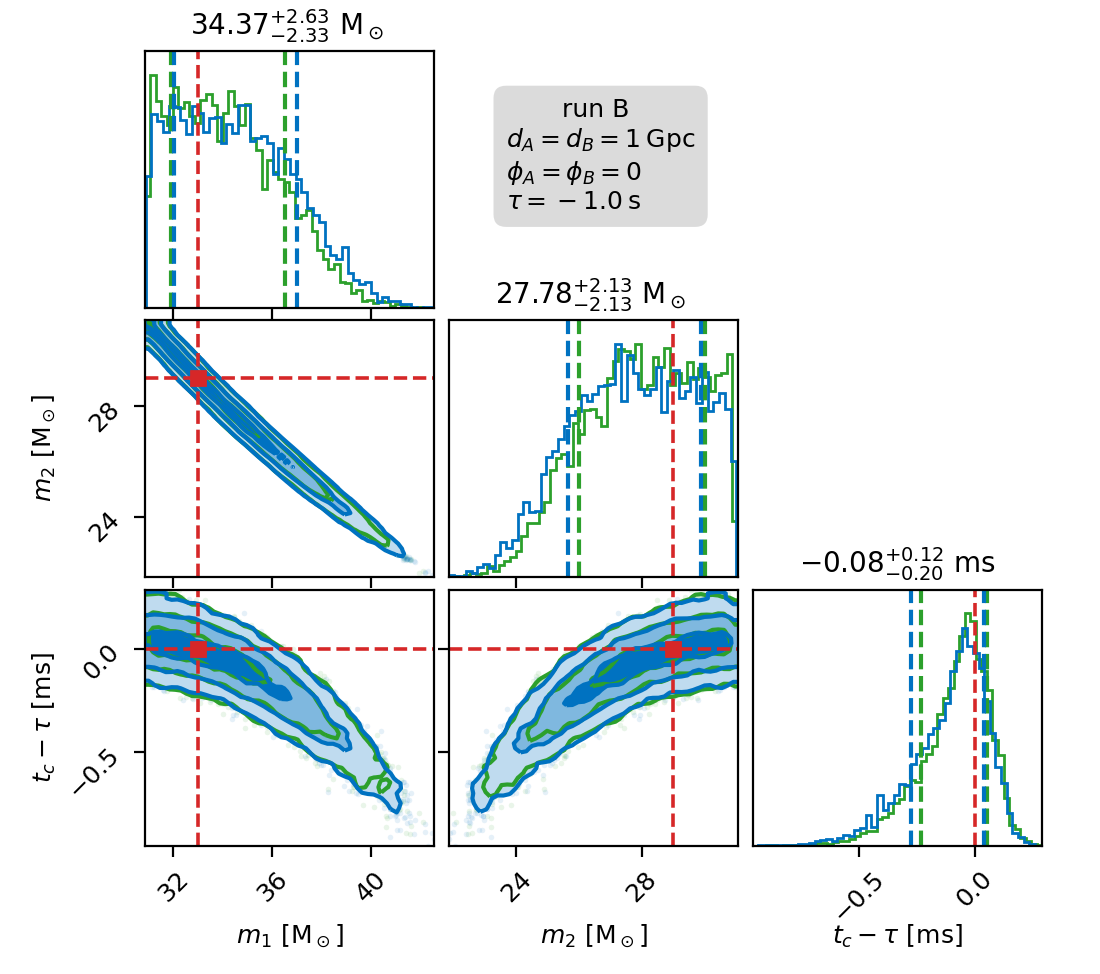} 
\includegraphics[width=0.48\textwidth]{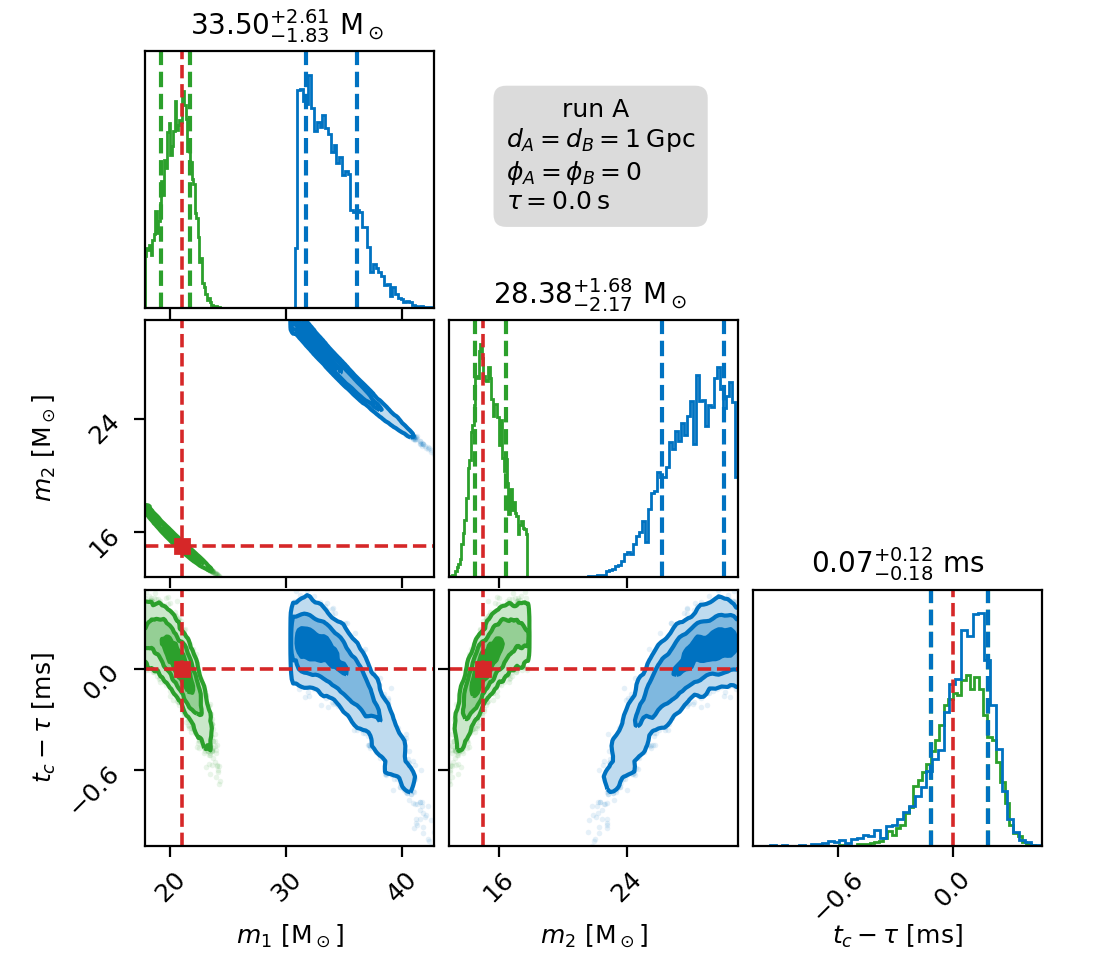}
\includegraphics[width=0.48\textwidth]{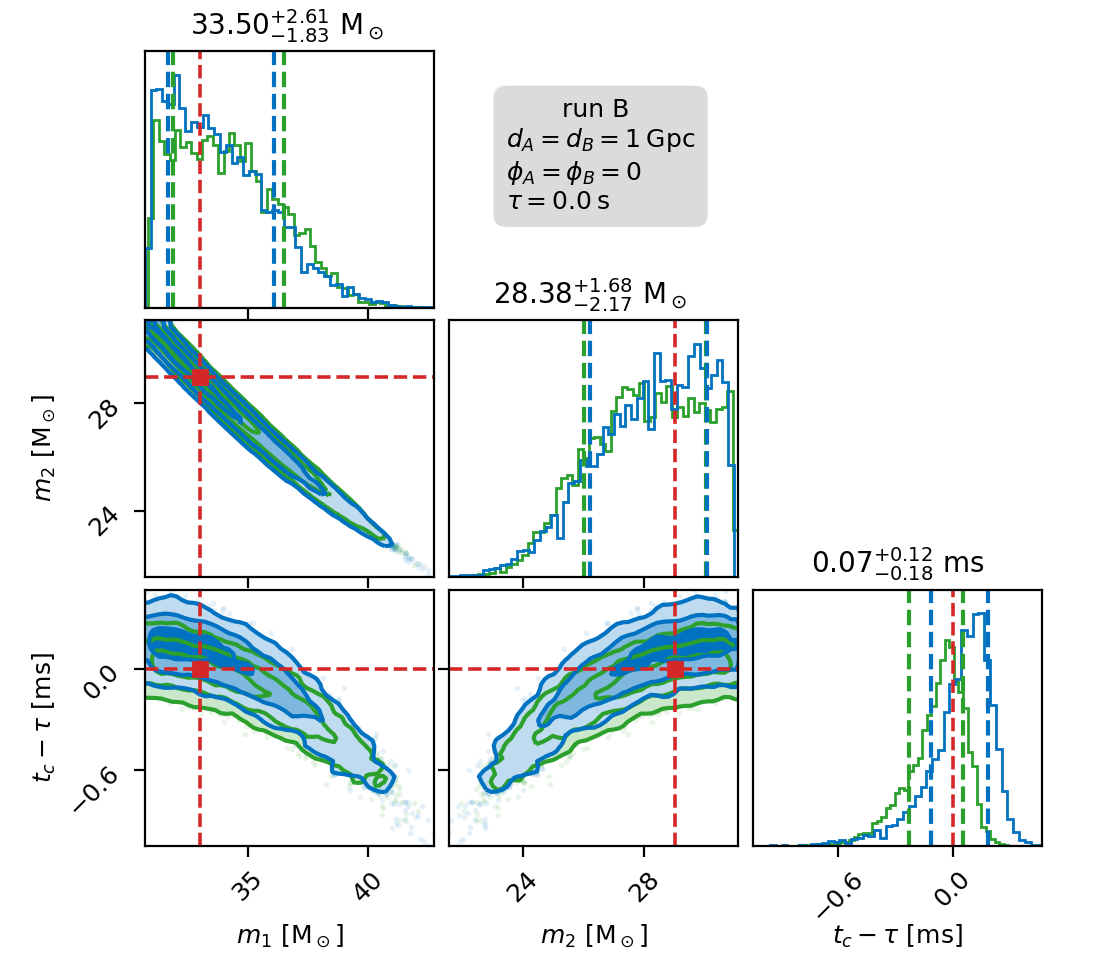} 
\end{center}
\vskip-0.5cm
\caption{Corner plots for two \textit{runs A} (left side) and two \textit{runs B} (right side); all the overlapping signals are created with the following choice of the parameters: $d_L^\mathrm{(A)}=d_L^\mathrm{(B)}= 1\,\mathrm{Gpc}$, $\phi_C^{(A)}=\phi_C^{(B)}=0$. The top row shows the case $\tau = -1.0\,\mathrm{s}$, while the bottom one shows $\tau=0.0\,\mathrm{s}$. The three parameters considered here are the two masses $m_1$ and $m_2$, and the coalescence time $t_C$. The true values of these parameters are highlighted with red dashed lines in the corner plots. The blue histograms refer to the actual runs, while the green ones are shown for comparison and they are obtained by injecting only one signal in the data. The dashed vertical lines represent the $1\sigma$ error on the parameters. On top of each panel, the median values (and the $1\sigma$ errors) of the parameters are shown.
}
\label{fig:corner}
\end{figure*}

In this section, we study the posterior distributions obtained for the different runs described in Sec. \ref{sec:parameters_run} and we compare them with the same results obtained when only a single signal is present in the data. This comparison allows us to assess the presence of biases created by the overlap of the signals. In this analysis, we focus on the results for the two masses $m_1$ and $m_2$ (which we can rewrite also as chirp mass $\mathcal{M}$ and mass ratio $q$), and for the coalescence time $t_C$. 

We start by plotting four different corner plots for specific values of the parameters (Fig. \ref{fig:corner}). In the top row, we show the posterior distributions for \textit{run A} (left panel) and \textit{run B} (right) for the following parameters: $d_L^{(A)} = d_L^{(B)} = 1\,\mathrm{Gpc}$; $\phi_C^{(A)}=\phi_C^{(B)}=0$; $\tau=-1.0 \mathrm{s}$. The blue contours represent the results obtained when the two signals are overlapping, while the green ones are the results for a run where only signal A (B) is present in the data. The agreement between these two posteriors (upper panels) is remarkably good, and biases, if any, are negligible. The recovered values of the parameters in the case of \textit{run A} (\textit{run B}) are perfectly compatible with the injected ones
$\lambda^{(A)}=\{m_1=21\,\msun, m_2=15\,\msun, t_C = 0.0\,\mathrm{s}\}$ ($\lambda^{(A)}=\{m_1=33\,\msun, m_2=29\,\msun, t_C = \tau\}$). This proves that using the current parameter inference methods to deal with overlapping signals is possible.

These results also imply that the posterior distribution for a run with wider priors would be (at least) bi-modal, as the two peaks identified by the two runs (corresponding to the true values of the parameters $\lambda_A$ and $\lambda_B$) with narrower priors would be preserved when the priors are extended coherently to a larger parameter space. However, as already mentioned in Sec. \ref{sec:priors}, when we try to extend the prior on the time shift $\tau$, we find that the sampling algorithm can identify only one peak in the posterior. This behavior is due to the fact that the heights of the two peaks differ by many orders of magnitude, since the peak of $\log \mathcal{L}$ scales as the SNR squared, and the SNRs for signal A and signal B are $\mathrm{SNR}^{(A)} = 27.1$ and $\mathrm{SNR}^{(B)} = 41.5$, respectively (see also Table \ref{tab:SNR}). Clearly, the sampling algorithm is not able to sample such a subdominant peak in the posterior. Thus, setting the appropriate prior on the coalescence time $t_C,$ as determined by the search pipeline, is critical in determining the parameters of both of the signals.

We note that a different approach could consist of imposing narrower priors on the two masses $m_1$ and $m_2$ (or, equivalently, on the chirp mass $\mathcal{M}$) in order to isolate one peak and exclude the other. This is also a viable alternative, provided that the information on the masses recovered from the detection pipeline is accurate enough to give effective constraints for the priors.
Ultimately, combining the information on the coalescence time with the one on the masses may be the best strategy in order to isolate the two peaks even when the two signals are coalescing very close to each other. It is, however, important to ascertain the extent to which such constraints can imposed by carrying out the detection problem on a large sample of injections and the accuracy with which detection pipelines are able to measure chirp mass. 

In fact, our approach fails when the two signals are overlapping within $100\,\mathrm{ms}$. In the bottom row of Figure \ref{fig:corner}, we show exactly this case: we take the same distances and phases as described above, but we impose a zero time shift between the two signals. Therefore, in this case the two runs \textit{run A} and \textit{run B} yield the exact same results (as both the priors and the likelihood are the same). As expected, only the louder signal (i.e., signal B) is correctly recovered, with the posterior distribution resembling very closely (although not perfectly matching) the one obtained in the single signal case. We conclude that, once again, the bias is negligible for \textit{run B}. As for signal A, the peak corresponding to $\lambda^{(A)}$ is completely neglected by our inference pipeline, and thus there is no way we can reconstruct the parameters of signal A correctly. This is an intrinsic limitation of our method: different inference prescriptions need to be devised in order to deal with the case of closely coalescing signals.

\subsubsection{Dependence on the luminosity distance} \label{sec:lum_distance}

We now analyze the results of the other runs, where we changed the time shift, luminosity distance, and phase of coalescence of the two signals (as described in Sec. \ref{sec:parameters_run}). The top row of Fig. \ref{fig:res_summary} shows the posterior distributions for the chirp mass $\mathcal{M}$, the mass ratio $q$, and the coalescence time $t_C$, for different combinations of luminosity distances $d_L^{(A)}, \,d_L^{(B)}$ and coalescence times $t_C$; the phase at coalescence of the two signals are set to $\phi_C^{(A)}=\phi_C^{(B)}=0$.

Posteriors are shown in the form of violin plots, and the results for a single injection are shown in light grey color for reference on the right side of each panel. In order to make the plots more accessible, we identify three different regions, highlighted by the shaded grey boxes. In the first region (no shade), biases are negligible: posteriors for \textit{run A} (\textit{run B}) closely resemble the ones obtained by injecting only one signal with the same luminosity distance $d^{(A)}$ ($d^{(B)}$). In this region, the presence of overlapping signal does not create any biases to parameter inference, and both signals can be recovered correctly. As expected from our Fisher analysis (Sec. \ref{sec:fisher}), this happens when the two signals are not coalescing too close to each other. In particular, we find that parameter inference is robust in the regions $t_C\lesssim-0.75\,\mathrm{s}$ and $t_C\gtrsim0.25\,\mathrm{s}$. Note that the asymmetry of these boundary values are expected, as the correlation between the two signals is not symmetric in $\tau$ (Fig. \ref{Fig:fisher_correlation_BBH}).

When $t_C = -0.5\,\mathrm{s}$ (light shaded region), we find that small biases (at the $1-2\sigma$ level) arise: this implies that the presence of the overlap causes a shift of the posterior peak in the parameter space, preventing the correct recovery of the true parameters $\lambda^{(A)}$ and $\lambda^{(B)}$ for the two signals. We note, however that these relatively small biases may not be a problem in reality, because the presence of the noise may create even larger biases, making these effects totally irrelevant. This depends, of course, on the noise level in the interferometer. 

It is also interesting to note that intensity of the biases vary with the relative strengths of the two signals (which are determined by the luminosity distances). In particular, biases for run A (run B) are smaller whenever signal A (signal B) is louder: this can be observed in the left (right) panel of Fig. \ref{fig:res_summary}, top row, as the posteriors colored in yellow and purple (red, blue, and green) are closer to the ones obtained in the case of a single signal.

Finally, in the last region (darker shade, $\tau=0.0\,\mathrm{s}$), two relevant effects take place at the same time. First, as already discussed, only the parameters of the louder signal can be recovered correctly. Since the results for \textit{run A} and \textit{run B} are perfectly identical (because they have identical settings), this implies that chirp masses are close to the one of signal A for the yellow and purple cases (as seen in the left panel), and close to the one of signal B in the red, blue, and green cases (as seen in the right one). On top of that, we note that even the louder signal seems to suffer from significant bias in the $\tau = 0.0\,\mathrm{s}$ case. This is again expected from our Fisher analysis (Fig. \ref{Fig:fisher_correlation_BBH}), as the correlations between the signals have a peak at zero time shift.

\subsubsection{Dependence on the phase}

In the bottom row of Fig. \ref{fig:res_summary}, we show the results for the runs with varying $\phi_C$. As described in Sec. \ref{sec:parameters_run}, we fix the luminosity distances to $d_L^\mathrm{(A)}=d_L^\mathrm{(B)}=1\,\mathrm{Gpc}$ and the phase at coalescence of signal A to $\phi_C^\mathrm{(A)}=0$, and vary $\phi_C^\mathrm{(B)}$ in the set $\phi_C^\mathrm{(B)}=\{0,\pi/3,2\pi/3$\}. Results are presented in the same form as the top row of Fig. \ref{fig:res_summary} (Sec. \ref{sec:lum_distance}). 

We find that the phase at coalescence plays an important role in determining whether inference suffers from significant biases or not. In particular, biases are greater for the two configurations $\phi_C^\mathrm{(B)} = \pi/3$ and $\phi_C^\mathrm{(B)} = 2\pi/3$. On top of that, they extend in a larger time span: the region where $2\sigma$ biases are present extend out to $\tau=-1.0\,\mathrm{s}$; for $\tau = -1.5\,\mathrm{s}$ and $\tau=-2.0\,\mathrm{s}$, they progressively become less severe until they become hardly detectable. Again, we find that biases arise only for negative values of the time shift $\tau$, in accordance with the asymmetric correlation amplitudes found in Fig. \ref{Fig:fisher_correlation_BBH}.

Overall, our Bayesian inference analysis confirms the results we found in Sec. \ref{sec:fisher} for BBH in Advanced LIGO detectors (Fig. \ref{Fig:fisher_correlation_BBH}, upper left panel). If the two BBH signals do not coalesce too close in the time domain (i.e., their coalescence times are separated by more than $\approx1.5\,\mathrm{s}$), then inference results are robust: two distinct peaks are present in the posterior, and they can be well-sampled if a suitable prior on the coalescence time is chosen. This is an interesting conclusion, as the vast majority of BBH signals are expected to belong to this category: from Fig. \ref{fig:overlapping signals}, we can estimate that only $0.01\%$ of the signals are expected to coalesce within $1\,\mathrm{s}$.

When the BBH signals do coalesce very close to each other ($|\tau|\lesssim 1\,\mathrm{s}$), though, biases at the $2-3\,\sigma$ level may arise, as the correlation between the two signals increases. These biases become even more dramatic as the time shift approaches to zero.

\section{Discussion and Outlook} 
\label{sec:conclusions}

We presented a Bayesian inference analysis in the case of overlapping gravitational waves signals. Our goal was to assess the capabilities of current Bayesian inference infrastructure to handle the non-trivial case of one or multiple overlaps taking place within a data segment. This problem is destined to play a major role in 3G detector planning, since the dramatic increase in sensitivity will result in a great number of signals coalescing within a few seconds. 

We started from a study based on the Fisher matrix formalism, in which we analyzed the correlation between two overlapping signals. In this way, we were able to determine whether in some regions of the parameter space the overlapping signals were strongly correlated with each other, thus preventing a distinct inference procedure for one signal at a time. We found that BNS signals are less strongly correlated, and that  their inference will likely be a problem only for coalescence times really close to each other (at the $10-100\,\mathrm{ms}$ level). BBHs, instead, suffer from the presence of a correlation starting from a much greater time shift $\tau$ (i.e., the difference between the two coalescence times). In particular, in the Advanced LIGO BBH scenario, correlations are significant up until $|\tau| \approx 2 \,\mathrm{s}$.

We investigated these issues further with a full Bayesian analysis of the two overlapping BBHs. The analysis used the \code{dynesty} sampling algorithm to describe the posterior distribution for the parameters considered. We showed that, in order to sample a single peak without worrying for the presence of the other one, a possible solution is to impose a narrow prior around the fiducial value (provided by the signal detection pipeline) of the coalescence time of the signal of interest. This procedure allows to isolate one single peak at a time, and works well in the configurations we explored. However, as the time shift approaches zero, isolating one single peak at a time is not possible, and within our framework we can recover only the parameters for the louder signal (i.e., the highest peak in the posterior). In our approach, we are implicitly assuming that signal detection will return the coalescence times of the two signals with an uncertainty lower then $\mathcal{O}(10-100)$ ms. This is a reasonable assumption, which, however, needs to be tested by a dedicated analysis dealing with BBH signals' recovery in the context of 3G detectors (see also \cite{Regimbau:2012ir, Meacher:2015rex} for the BNS case).

We also studied the emergence of biases in the overlapping signals scenarios considered, by varying some key parameters of the two signals such as their coalescence time, coalescence phase, and luminosity distance. We found that significant biases (at the $2-3\sigma$ level) arise in the range $-1\,\mathrm{s}<\tau<0\,\mathrm{s}$, and that these biases are caused primarily by the relative phase of the two signals and only marginally by the relative difference of the SNRs. As suggested by our Fisher analysis (Fig. \ref{Fig:fisher_correlation_BBH}, upper left panel), these biases tend to become minor for $\tau<-1.5\,\mathrm{s}$ and $\tau>0\,\mathrm{s}$.

Dealing with these biases needs a different approach that we did not attempt in this work. One possible solution is to broaden the parameter space searching for multiple signals in the same Bayesian inference run. This is the approach that previous works have shown to be feasible in the context of LISA data analysis (e.g., see \cite{Cornish:2005qw,Littenberg:2020bxy}). Such approach could significantly increase the computational costs of the Bayesian algorithms; however, this is compensated by the fact that - as suggested here - novel algorithms may be needed only for closely-coalescing signals, that are a very small minority of the total number of signals expected in 3G detectors. Using current estimates for the BBH rates in future detectors, we find that signals coalescing within $1\,\mathrm{s}$ are expected to be at most hundreds per year. 

Another possible solution to the biases would be to create an iterative procedure where one hierarchically determines the parameters of louder signals (as inferred from search algorithms) and subtracts them from the data before analysing weaker ones \cite{Cutler:2005qq, Sachdev:2020bkk}. Currently, it is unclear which approach will perform better in the context of 3G detectors, and further work is needed to gauge the potential of both approaches.

In our exploratory study we did not deal with the consequences of varying the mass parameters of the two signals, nor did we include in our analysis other source parameters such as companion spins, the position of the source in the sky and the orientation of the binary relative to the detector frame. The SNR range explored in our study (20-100; see also Tab. \ref{tab:SNR}) is also limited compared to the range expected to be covered by 3G detectors \cite{Punturo:2010zz, Reitze:2019dyk, Borhanian:2022czq}. 
In particular, when overlapping signals arrive from different positions in the sky then they would, in general, have different coalescence times in different detectors, which might help to isolate one of the peaks better \cite{Christian:2018vsi}. The inclusion of spins, on the other hand, introduces new physics in the formation of these overlapping signals such as spin precession, and may introduce another layer of complexity in the parameter inference problem \cite{Fairhurst:2019vut}. These and related problems will be explored in a future study. 

\section*{Acknowledgments} 
We thank Anuradha Samajdar, Justin Janquart, Chris Van Den Broeck and Tim Dietrich for sharing and discussing their results on a similar study \cite{Samajdar:2021egv}.  
We are indebted to useful comments by Rory Smith and Salvatore Vitale.
EP is grateful to Walter Del Pozzo for helpful suggestions.
EP was supported by INFN in the framework of the 2019 NSF-INFN Summer exchange program. 
SS is supported by the Eberly Postdoctoral Fellowship of Penn State. 
BSS is supported in part by NSF Grant No. PHY-1836779, PHY-2012083, and AST-2006384.
The authors are grateful for computational resources provided by the LIGO-Caltech Computing Cluster.
This paper has the LIGO document number P2100044.

\bibliographystyle{JHEP}
\bibliography{references}

\end{document}